\documentclass[a4paper,twosides]{article}
\usepackage{multicol,multirow,graphics,fancyhdr,epstopdf}
\usepackage{amsmath,amsfonts,amssymb,bm,upgreek,mathrsfs,ccmap,mathcomp}

\usepackage{verbatim}
\usepackage[pagewise,switch,columnwise]{lineno}
\usepackage[compress,nospace]{cite}
\usepackage[dvipsnames]{xcolor}
\usepackage{CPL-2022}

\newcommand{\cplyear}{2023} \newcommand{\cplvol}{40}
 \newcommand{\cplpagenumber}{122101}
 \newcommand{\cplpage}{\cplpagenumber-\thepage}

\usepackage{tikz,xcolor,hyperref}
\definecolor{lime}{HTML}{A6CE39}
\DeclareRobustCommand{\orcidicon}{
	\begin{tikzpicture}
	\draw[lime, fill=lime] (0,0) 
	circle [radius=0.16] 
	node[white] {{\fontfamily{qag}\selectfont \tiny ID}};
	\draw[white, fill=white] (-0.0625,0.095) 
	circle [radius=0.007];
	\end{tikzpicture}
	\hspace{-2mm}}

\foreach \x in {A, ..., Z}{%
	\expandafter\xdef\csname orcid\x\endcsname{\noexpand\href{https://orcid.org/\csname orcidauthor\x\endcsname}{\noexpand\orcidicon}}
	}

\begin{document}

\vspace* {-4mm} \begin{center}
\large\bf{\boldmath{Phase Transition Study meets Machine Learning}}
\footnotetext{\hspace*{-5.4mm}$^{*}$Email: 
mayugang@fudan.edu.cn; lgpang@mail.ccnu.edu.cn; rui.wang@ct.infn.it; zhou@fias.uni-frankfurt.de

\noindent\copyright\,{\cplyear}
\href{http://www.cps-net.org.cn}{Chinese Physical Society} and
\href{http://www.iop.org}{IOP Publishing Ltd}}
\\[5mm]
\normalsize \rm{Yu-Gang Ma$^{1,2,*}$, Long-Gang Pang$^{3,*}$, Rui Wang$^{1,4,5,*}$, Kai Zhou$^{6,*}$}
\\[3mm]\small\sl $^{1}$ Key Laboratory of Nuclear Physics and Ion-beam Application~(MOE), and Institute of Modern Physics, Fudan University, Shanghai $200433$, China
\\[3mm]\small\sl $^{2}$Shanghai Research Center for Theoretical Nuclear Physics, NSFC and Fudan University, Shanghai 200438, China
\\[3mm]\small\sl $^{3}$ Institute of Particle Physics and Key Laboratory of Quark and Lepton Physics (MOE), Central China Normal University, Wuhan, 430079, China
\\[3mm]\small\sl $^{4}$ Shanghai Institute of Applied Physics, Chinese Academy of Sciences, Shanghai $201800$, China
\\[3mm]\small\sl $^{5}$ INFN Sezione di Catania, 95123 Catania, Italy
\\[3mm]\small\sl $^{6}$
Frankfurt Institute for Advanced Studies (FIAS), D-60438 Frankfurt am Main, Germany

\end{center}
\vskip 1.5mm

\small{\narrower In recent years, machine learning (ML) techniques have emerged as powerful tools in studying many-body complex systems, encompassing phase transitions in various domains of physics.
This mini-review provides a concise yet comprehensive examination of the advancements achieved in applying ML for investigating phase transitions, with a primary emphasis on those involved in nuclear matter studies. 
\par}\vskip 3mm
\normalsize\noindent{\narrower{PACS: 21.65.-f, 21.65.Mn, 21.65.Qr, 07.05.Mh, 05.70.Fh}}\\
\noindent{\narrower{DOI: \href{http://dx.doi.org/10.1088/0256-307X/40/12/122101}{10.1088/0256-307X/40/12/122101}}

\par}\vskip 5mm

As a foundational concept in the exploration of many-body physical systems, phase transitions encapsulate phenomena as commonplace as the freezing of water into ice, and as esoteric as the postulated quark-gluon plasma (QGP) to hadronic matter transition believed to have occurred in the early universe or during heavy-ion collisions. Phase transitions across these various physical domains share fundamental similarities, primarily characterized by a drastic change in the state of matter due to thermal fluctuations and/or quantum effects. The universal features such as critical exponents and symmetry breakings forge a connection among disparate systems, enabling the transference of insights and methods across disciplinary boundaries. Nevertheless, each physical system brings with it unique challenges and intricacies. Specifically, the study of phase transitions of nuclear matter governed by quantum chromodynamics (QCD) is crucial to our understanding of the nature of strong interactions and has burgeoned into a pivotal research area \ucite{Aarts:2023vsf,Bzdak:2019pkr}. This includes explorations of, e.g., nuclear liquid-gas phase transition \ucite{PocPRL75,MYGPRL83,NatPRL89,NatPRC65,MYGPRC71,Liu,PPNP,DengXG}, as well as the phase structure of hot and dense QCD matter \ucite{Fukushima:2010bq,Saiz,LiX}. However, the non-perturbative nature of QCD especially at low energies makes such information of phase transition particularly elusive, often surpassing the capabilities of traditional theoretical methods. In general, studying phase transitions in complex systems, especially those with many degrees of freedom, usually poses significant computational hurdles. In this context, machine learning (ML) techniques~\ucite{He:2023zin,NT_LiF,SCPMA_He,SSPMA-2021-0299,NST_Shang,SCPMA_Liang,CPL_Zhang,NST_Li} offer novel and promising pathways for its investigations. 

This short review aims to elucidate the synergies between ML \ucite{SCPMA_HeLS,NST_Xie,CPL_Bai,NST_Ming,CPL_Gao,NT_Gao,NT_Niu} and phase transition studies, and how these interdisciplinary collaborations are paving the way for new insights in physics. We will delve into recent breakthroughs where ML methods, including supervised and unsupervised learning, have been applied to phase transition studies of nuclear matter, enabling a finer characterization of the QCD phase diagram and a deeper understanding of the nuclear matter properties. We will also discuss the challenges and future directions in this exciting and rapidly evolving field.

{\textbf{Phase transition in condensed matter physics}}

\begin{figure*}[htb]
\centering
\includegraphics[width=16cm]{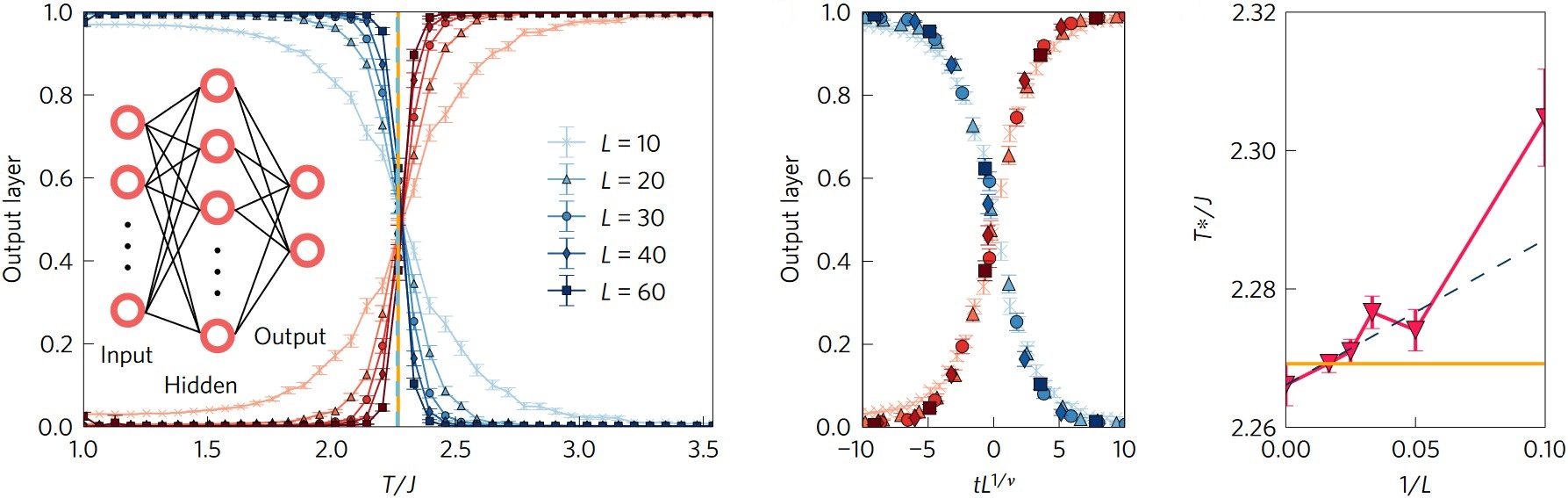}
\put(-460,135){\bfseries (a)}
\put(-235,135){\bfseries (b)}
\put(-115,135){\bfseries (c)}
\caption{\small (Color online) 
 (a) The output of the neural network averaged over a test set as a function of $T/J$.
 The input of the neural network is the 2-dimensional spin configuration of the square-lattice ferromagnetic Ising model generated by Monte Carlo simulations.
 Different system sizes $L$ are used. 
 The inset shows the sketch of the neural network in use.
 (b) Re-scaling of the output in (a) as $T/J$ $\rightarrow$ $tL^{1/\nu}$ with $t$ $=$ $(T-T_c)/J$.
 With proper $\nu$, these lines can collapse to each other.
(c) Plot of the finite-size scaling of the crossing temperature $T^*/J$ (down triangles).
The horizontal line represents the theoretical critical temperature in the thermodynamic limit $T_c/J$ $=$ $2/ln(1 + \sqrt{2})$.
Picture from Ref.~\cite{CarNtP13}.}
\label{F:Ising}
\end{figure*}

(1) {\it{Supervised learning and Ising model}} :
The first application of ML techniques to phase transitions emerged in the field of condensed matter physics~\ucite{CarNtP13}.
ML techniques are designed to characterize complex datasets, which makes these techniques rather suitable for recognizing and classifying the phases of matter. Here we take the example of employing the fully connected neural network [the inset in Fig~\ref{F:Ising}(a)] to the square-lattice ferromagnetic Ising model ($H$ $=$ $-J\sum_{ij}\sigma_i^z\sigma_j^z$, with $J$ being the interaction and $\sigma_i^z$ $=$ $\pm1$ being the spin).
The data set used for training and testing the neural network consists of 2-dimensional spin configurations sampled by Monte Carlo simulations for different temperatures $T$.
The neural network was trained to classify the ferromagnetic configurations (below critical temperature $T_c$) and the paramagnetic configurations (above $T_c$).
Figure~\ref{F:Ising}(a) shows the output of the neural network averaged over a test set as a function of $T/J$ for different system sizes represented by $L$.
Clearly, the trained neural network could respond differently to spin configurations with low temperatures and those with high temperatures.
The temperature at which the two outputs are equal to each other, namely the crossing temperature, $T^*$, can be viewed as the $T_c$ obtained by the neural network for the particular $L$.
In the language of renormalization, one can re-scale those $T/J$ dependence of the output, i.e., $T/J$ $\rightarrow$ $tL^{1/\nu}$ with $t$ $=$ $(T-T_c)/J$, to obtain the critical exponent $\nu$ $\approx$ $1.0\pm0.2$ [Fig.~\ref{F:Ising}(b)].
With the $1/L$ dependence of $T^*/J$ shown in Fig.~\ref{F:Ising}(c), by taking the limit $L$ $\rightarrow$ $\infty$, one can obtain the crossing temperature from the neural network $T^*/J$ $=$ $2.266\pm0.002$, which estimates quite accurately the physical critical temperature of the square-lattice Ising model.
The latter can be obtained theoretically in the thermodynamic limit, i.e., $T_c/J$ $=$ $2/ln(1 + \sqrt{2})$.

Neural networks are even able to generalize the tasks beyond their original design.
For the trained neural network in the case of the square-lattice Ising model, when fed with the spin configurations obtained by Monte Carlo simulations of the triangular-lattice ferromagnetic Ising Hamiltonian, the neural network successfully estimates the critical temperature $T_c/J$ $=$ $3.65\pm0.01$ and $\nu$ $\approx$ $1.0\pm0.3$, which are consistent with the theoretical values. 
Using a more complicated convolutional neural network, it is even possible to encode basic information about unconventional phases, such as those present in the square-ice model and the Ising lattice gauge theory, as well as the Anderson localized phases~\ucite{CarNtP13}.

(2) {\it{Semi-supervised and unsupervised learning}} :
Besides the supervised learning technique, in Ref.~\cite{NieNtP13}, a confusion scheme combining supervised and unsupervised learning has been proposed to classify the phases of matter.
The main idea of the confusion scheme is as follows.
First, classify the data set with a proposed critical point $u'_c$.
Then, supervised learning, such as the fully connected network, is applied to the data set to obtain the test accuracy of the trained neural network for each $u'_c$.
The test accuracy, as a function of $u'_c$, will exhibit a $W$-shape, and the true critical point corresponds to the middle peak of the $W$-shape.
Since in this confusion scheme, one does not require the knowledge of the order parameters, the topological content of the phases, or any other specifics of the transition in question, it thus provides a generic tool for identifying unexplored phase transitions.

Another well-known ML technique for studying the phase transitions in condensed matter physics is the auto-encoder method~\ucite{WetPRE96,HWPRE95}, which is a pure unsupervised learning technique.
The auto-encoder neural network has a mirror structure.
The encoder part encodes the input layer into one or more latent variables, and the decoder part decodes the latent variables to the output layer.
The neural network is trained to best recover the encoded information, i.e., the network is trained to minimize the difference between the input layer and the output layer.
The auto-encoder method has been shown to be a very effective algorithm for examining phase transitions for several typical models in condensed matter physics, such as the ferromagnetic Ising model and the XY model~\ucite{WetPRE96,HWPRE95}.

{\textbf{Nuclear temperature}}

(1) {\it{Introduction}} :
One of the difficulties in studying the nuclear liquid-gas phase transition of nuclear matter arises mainly from the preparation of a finite-temperature nuclear system and the determination of its temperature.
Heavy-ion collisions (HICs) provide a possible venue for studying the finite temperature properties of nuclear matter.
During the collision reaction, a transient excited system is formed, which can generally be regarded as a (near) equilibrium state, since the evolution of its constituent nucleons is sufficiently short compared to the global evolution.
Its temperature can be obtained from, e.g., energy spectra by moving source fitting, excited state populations, (double) isotope ratios, or quadruple momentum fluctuations. For a brief review, see Ref.~\ucite{EPJA_JBN}.
For a reliable thermometer of heavy-ion collisions, we require it to be insensitive to both the collective effects and the secondary decay of unstable nuclei after the system disintegrates, which is generally difficult to achieve.
Moreover, because of the difficulty in verifying the accuracy of the apparent temperature ($T_{app}$) obtained by these thermometers, it is not a trivial task to propose different ways of determining $T_{app}$ and thus provide more opportunities for cross-checking.

(2) {\it{DNN to determine Nuclear Temperature}} :
Using machine learning techniques, the charge multiplicity distribution can be used to determine $T_{app}$ of HICs at intermediate to low energies \ucite{SongYD_PLB}.
Usually, the fragment charge distributions $M_{\rm c}(Z_{\rm cf})$ show typical changes of the hot nuclei disassembly mechanism with temperature, i.e., from evaporation mechanism at a lower temperature, 
to multifragmentation at intermediate temperature,
till vaporization at higher temperatures.

A relation between the source temperature of the model $T^{model}$ and $M_{\rm c}(Z_{\rm cf})$ can then be established by a  deep neural network (DNN) that transforms the complex relation into a nonlinear map through its neurons.
This relation can be used to determine $T_{app}$ of a particular transient state during HICs at intermediate-to-low energies.
With the final state charge multiplicity distribution simulated with an isospin-dependent quantum molecular dynamics  (IQMD), its apparent temperature was obtained by the trained DNN.

With the $T_{app}$ determined by DNN, the caloric curve, i.e., $T_{app}$ as a function of the excitation energy per nucleon ($E^*/A$) of the HICs, was also examined.
Traditionally, the characteristic behavior of the caloric curve is explained by the fact that, as $E^*/A$  increases, the system is driven into a spinodal region, in which part of $E^*/A$ begins to transfer to latent heat. From the simulation, the extracted $T_{app}$ at the maximum of the specific heat capacity of the system $\tilde{c}$ $\equiv$ $d(E^*/A)/dT_{app}$, which is called the limiting temperature $T_{\rm lim}$,  
follows the general trend of Natowitz's limiting-temperature dependence on the system size~\ucite{NatPRC65}, and thus indicates the validity of determining $T_{app}$ through charge multiplicity distribution with the help of ML techniques.

{\textbf{Nuclear liquid gas phase transition}}

(1) {\it{Introduction}} :
As mentioned above, machine-learning techniques can be used to study nuclear temperature and then the caloric curve, which is of particular interest in reaction dynamics. 
A lot of probes by analyzing sophisticatedly the information of the reaction products have been proposed to recognize the liquid-gas phase transition of nuclei~\ucite{PPNP,PocPRL75,MYGPRL83,NatPRL89,NatPRC65,MYGPRC71,DengXG,Liu}.
Recently, a specific effect from initial-state $\alpha$-clustering configurations \cite{Ma:2022dbh,PLB_Wang,PRC_Wang,shicz2,Ma_NuclTech} on the nuclear liquid-gas phase transition was also discussed \cite{PRC_Cao}.
Since the nucleus is an uncontrollable system, its liquid-gas phase transition is realized by tracing the effect of the spinodal instability on the reaction dynamics, e.g., by measuring the properties of the intermediate mass fragments~(with charge number greater than $3$).

(2) {\it{DL Confusion Sheme for Nuclear Liquid-Gas Phase Transition}} :
In a recent work, the averaged charge multiplicity distribution $\langle M_{\rm c} \rangle(Z)$ of the quasi-projectile (QP) fragments in the reactions of $^{40}{Ar}$ on $^{27}{Al}$ and $^{48}{Ti}$ at $47~\rm MeV/nucleon$ has been used to study the nuclear liquid-gas phase transition by the autoencoder method and a confusion scheme \ucite{WangR_PRR}.
The QP fragments are supposed to come from the excited projectile nucleus, which can largely avoid the effect of dynamical evolution.
They can be obtained by a three-source (i.e., a QP source, an intermediate velocity source, and a quasi-target source) reconstruction method~\ucite{MYGPRC71}.

\begin{figure*}[htb]
\centering
\includegraphics[width=16cm]{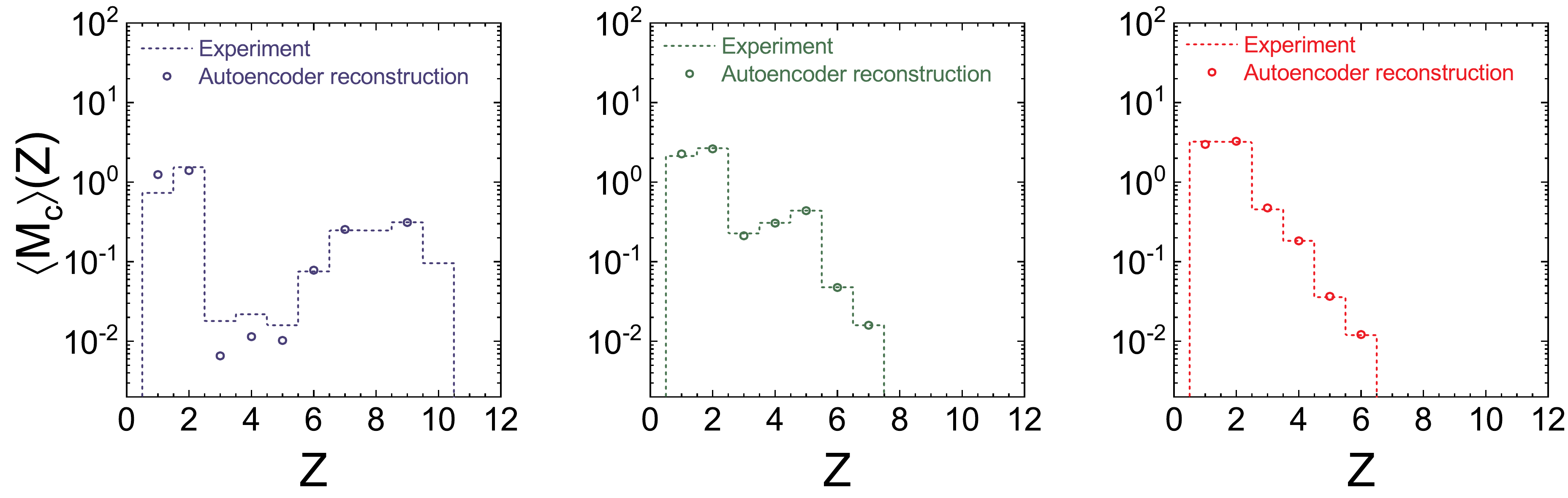}
\put(-380,105){\bfseries Evaporation}
\put(-235,105){\bfseries Mult-}
\put(-235,95){\bfseries fragmentation}
\put(-80,100){\bfseries Vaporization}
\put(-452,136){\bfseries (a)}
\put(-299,136){\bfseries (b)}
\put(-146,136){\bfseries (c)}
\caption{\small (Color online)  The averaged charge multiplicity distribution $\langle M_{\rm c} \rangle(Z)$ of the QP fragments.
The average is taken for different $E^*/A$ bins, left panel for low excitation~($0.9~\rm MeV$ - $2.8~\rm MeV$), middle panel for intermediate excitation~($5.3~\rm MeV$ - $5.4~\rm MeV$), and right panel for high excitation~($8.1~\rm MeV$ - $13.0~\rm MeV$).
The dashed curves represent $\langle M_{\rm c}\rangle(Z)$ from the NIMROD experiment, while the circles from the autoencoder network reconstruction $\langle M_{\rm c}'\rangle(Z)$.
Each $E^*/A$ bin contains $500$ test events. Picture from Ref. \cite{WangR_PRR}.}
\label{F:CM}
\end{figure*}

\begin{figure*}[b]
	\centering
	\includegraphics[width=.6\columnwidth]{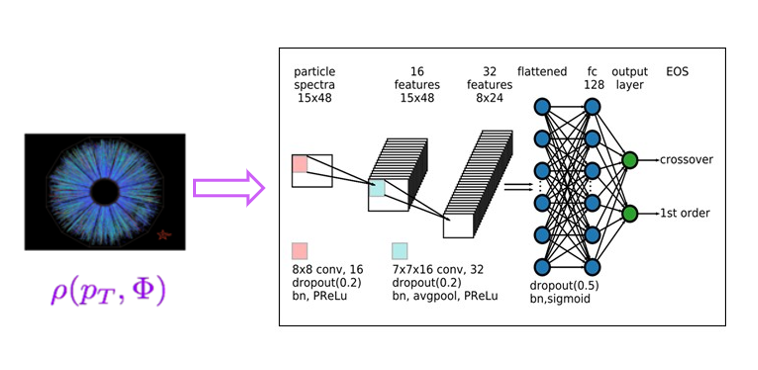} 
	\caption{Flowchart of QCD phase transition identification with heavy ion collision particle spectra.}
	\label{pt_hic}
\end{figure*}

\begin{figure*}[b]
	\centering
	\includegraphics[width=.38\columnwidth]{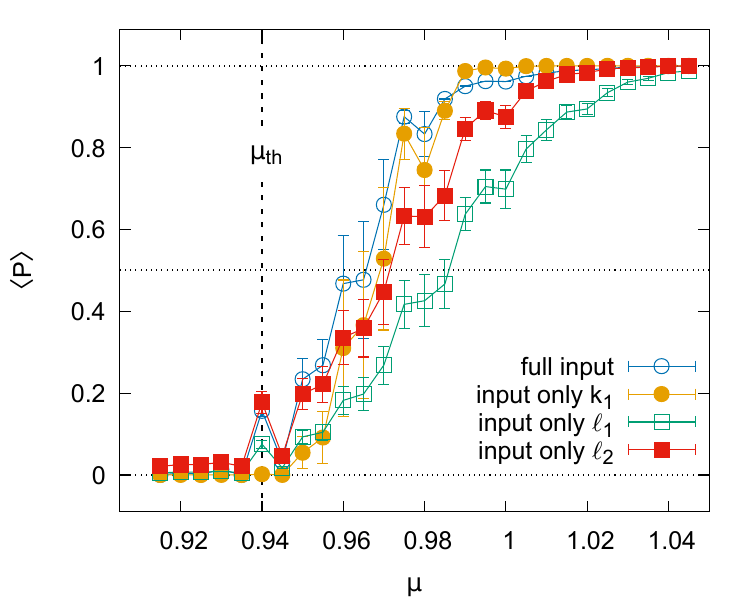}
        \includegraphics[width=.42\columnwidth]{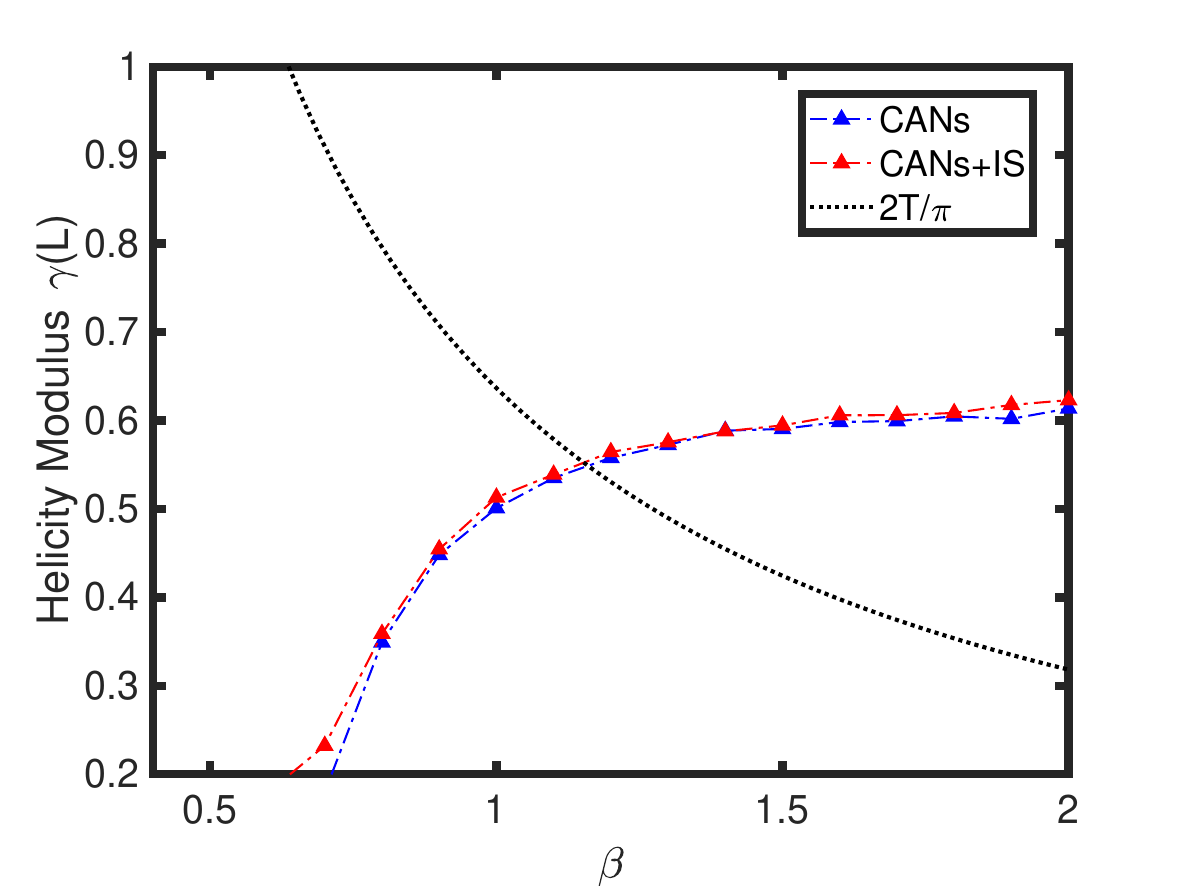}
	\caption{(left) The average condensation $\langle{P(\phi)}\rangle$ of as a function of the chemical potential $\mu$ using full and restricted field inputs. Picture from Ref.~\cite{Zhou:2018ill}. (right) The helicity modulus of 2D XY model from CANs and CANs-IS cross with the curve $2k_BT/\pi$ to derive the transition point, $\beta_{\rm c}\simeq 1.1796$. Picture from Ref.~\cite{Wang:2020hji}.}
	\label{probcond}
\end{figure*}

The event-by-event charge-weighted charge multiplicity distribution of QP fragments $ZM_{\rm c}(Z)$ from the experiment are used as the input to train the autoencoder network.
The network consists of two main parts, the encoder part encodes the input event-by-event $ZM_{\rm c}(Z)$ into a \emph{latent variable}, and the decoder part decodes the latent variable into $ZM'_{\rm c}(Z)$.
The neural network is trained to recover the encoded information as best as possible, i.e., the network is trained to minimize the difference between $ZM_{\rm c}(Z)$ and $ZM'_{\rm c}(Z)$.
Fig.~\ref{F:CM} shows the averaged $M_{\rm c}(Z)$ in three typical $E^*/A$ bins (dashed lines).
For the test QP events, the reconstructed $M_{\rm c}'(Z)$ are averaged and compared with the original $M_{\rm c}(Z)$ in Fig.~\ref{F:CM}.
Once the autoencoder network is trained, each QP event is mapped to the latent variable.
The latent variable as a function of $T_{app}$ and $E^*/A$ shows a sigmoid pattern, indicating that the trained autoencoder network treats the low and high temperature~(or low and high excitation energy) regions differently.
The area in the midst of the two phases represents those liquid-gas coexistence events that enter the spinodal region and are affected by the spinodal instability.

The confusion scheme mentioned above has been adopted to obtain the limiting temperature of the nuclear liquid-gas phase transition.
The neural network is trained with the event-by-event charge-weighted charge multiplicity distribution of QP fragments $ZM_{\rm c}(Z)$ that was labeled according to a proposed limiting temperature $T_{\rm lim}'$.
The phase transition properties can be deduced from the performance curve, i.e., the total test accuracy as a function of the proposed $T_{\rm lim}'$, of the neural network.
The total test accuracy reaches its minimum at $T'_{\rm lim}$ $\approx$ $T_{\rm lim}$.
The limiting temperature through the confusion scheme is $9.24\pm0.04~\rm MeV$, which is consistent with the $9.0\pm0.4~\rm MeV$ obtained from the traditional analysis of caloric curve~\ucite{WadaPRC99}.

{\textbf{Transition identification in high energy HICs}}

(1) {\it{Introduction}} :
It's conjectured that at high temperature and (or) density, the QCD matter may experience deconfinement transition where normal nuclear matter or hadron resonance gas would turn to a new state of matter, called quark-gluon plasma (QGP). There are two types of deconfinement phase transitions relevant to QCD: one is crossover transition at vanishing or relatively small baryon chemical potentials, and the other is first-order phase transition at high baryon chemical potentials. The point at which these two types of transitions meet in the QCD phase diagram is named the critical point~\ucite{NT_WuYF,NT_LuoXF,NT_SunKJ,NT_ChenQ}. The first principle lattice QCD study currently only has an achievable prediction for the QCD phase diagram in high-temperature and low-baryon chemical potential regions, while in high-density regions the calculation is hindered by the fermionic sign problem. As the only way on the earth to heat and compress nuclear matter for creating and studying QCD phase transition, HICs involve a complex and superfast evolving process, where though interesting physics including color deconfinement may indeed have occurred in the early stage during the collision, the excited fireball will quickly expand with the temperature drops soon to confinement region, thus experimentally one can only resolve the finally emitted hadrons from HICs.

(2) {\it{CNN based Identification of QCD Transition}} :
With merely the final state pions' spectra, a deep learning-based EoS-meter was proposed to identify the QCD transition type that happened in the early stage of HICs~\ucite{Pang:2016vdc}. At the level of relativistic hydrodynamics, binary classification of QCD transition -- crossover or first order, is performed by a convolutional neural network (CNN) with final state pion's spectra as input. See Fig.~\ref{pt_hic} for a demonstration. As a supervised learning task, the dataset can be generated from the well-developed hydrodynamic simulation for HICs, with the labels provided by the implemented two classes of EoSs incorporating either crossover or first-order transition type in each generated collision event. The trained CNN shows above $95\%$ event-by-event classification accuracy, which is also robust against different initial fluctuations and shear viscosities, indicating that information about QCD transition that happened in the early time of HICs is not fully washed out by the evolution and still encoded in the final state, and can be decoded with CNN to help identify the phase transition with experimental measurements. 

(3) {\it{Extend the Classifier to Realistic}} :
This strategy was further extended to more realistic cases, e.g., including the afterburner hadronic cascade evolution via UrQMD simulation into the simulation in the transition identification~\ucite{Du:2019civ}, where the particlization, hadronic rescattering, and more stochasticity appeared inside the dynamics. The classification accuracy for the QCD transition type from CNN indeed decreased on an event-by-event basis, however, it was demonstrated that with fine-centrality-bin average the classification can be well achievable by CNN; the non-equilibrium phase transition scenario was also discussed via introducing spinodal decomposition induced baryon clumping into the simulation for the classification task~\ucite{Steinheimer:2019iso}. While in position space the CNN can easily distinguish such non-equilibrium transition type from normal Maxwell reconstructed transition inside the EoS, it strives when only momentum space spectrum is available, though PointNet based model could give slight enhancement, indicating that the characteristic of the two classes exists on ensemble level; Moreover, the analysis was recently pushed to use directly the experimental detector readout, where PointNet based model was accordingly explored to identify the QCD transition type for CBM experiment~\ucite{OmanaKuttan:2020btb}.

{\textbf{EoS study from HICs}}

(1) {\it{Introduction}} :
In the context of HICs, the equation of state (EoS) of nuclear matter, which essentially encodes the QCD phase diagram and transition information, has long been one of the holy grails of both theoretical and experimental research. From a theoretical perspective, the EoS is a crucial input in our physical modeling for HICs, directly influencing the dynamical evolution of the collisions. As for the dense isospin-asymmetric nuclear matter, which is expected to be produced in low to intermediate-energy HICs, its density-dependent EoS is currently not understood and not accessible yet to first principle lattice QCD calculations. 

(2) {\it{CNN Constrain Symmetry Energy}} :
One way to generally characterize dense nuclear matter is using nuclear empirical parameters defined from series expansion around the isospin asymmetry~\ucite{NST_An,SCPMA_LI,NST_Xie,NST_Xu}. Among them, nuclear symmetry energy is critical in encoding the energy cost to make the matter more neutron-rich. It's thus intriguing to get effective constraints over the nuclear symmetry energy with measurements in HICs, which, however, didn't yet give sensitive inference to the high-density part with final state particle distribution. Applying CNN on the final state proton spectra~\ucite{Wang:2021xbb}, it was demonstrated that the types of nuclear symmetry energy can be reasonably classified, and the slope parameter can be well identified under regression training. 

(3) {\it{Bayesian Inference of Dense QCD matter EoS}} :
Recently, for the first time, the Bayesian inference was performed to constrain the density dependence of the nuclear matter EoS using observables from HICs in the beam energy range $\sqrt{s_{NN}}=2 - 10$ GeV\ucite{OmanaKuttan:2022aml}. Specifically, the elliptic flow and mean transverse kinetic energy of protons were used in the analysis, which gives tight constraints on the EoS up to 4 times the nuclear saturation density. As a robustness check for predictability, the extracted EoS generates good agreement with other observables (the directed flow and differential elliptic flow) that are not used in the Bayesian analysis and also showed comparable speed of sound constrained from Neutron Star observations~\ucite{Soma:2022vbb}.

{\textbf{Machine learning phase transition in lattice QFT}}

(1) {\it{Introduction}} :
Besides experimental study, phase transition can also be studied by lattice QFT, where configurations of the many-body system can be simulated via Monte Carlo techniques to further evaluate different observables and partition functions.

(2) {\it{CNN and GAN for Scalar Field Study}} :
Phase identification for the 1+1-d complex scalar field was explored~\ucite{Zhou:2018ill} based on microscopic field configurations. For the considered system, the ``Silver blaze'' behavior is expected, where the particle density will be suppressed at low chemical potential until some threshold $\mu_{th}$ it will then increase considerably and enter a condensation region.  A convolutional neural network was trained for identifying the condensation phase with a training set consisting of only two ensembles of field configurations: one well above and one well below the transition chemical potential $\mu_{th}$. After training, the testing was performed on ensembles of configurations at different chemical potential values in between the two training chemical potential values. It was found that the ensemble average of the network output serves as an accurate phase classifier, $\langle P(\phi)\rangle$, with its non-vanishing point along with increasing chemical potential precisely indicating the transition point (see the left panel of Fig.~\ref{probcond}). Physical observables like the order parameter of the system--particle density $n$, and the squared field $|\phi^2|$ were also shown to be capable of being regressed well with the CNN. Ref.~\cite{Zhou:2018ill} further discussed using generative adversarial networks (GAN) to provide a fast, efficient sampler for independent field configurations' generation, which is shown to be capable of well capturing the underlying statistical distribution of configurations for the system and thus also can get the partition function learned essentially.

(3) {\it{Variational Autoregressive Generation of Field Configurations}} :
Deep generative models in machine learning provide an efficient way to study the phase structure of statistical many-body systems, which could replace or assist the conventional Markov Chain Monte Carlo simulation in related studies. Recently, a variational approach with an autoregressive network (VAN) is devised to solve Ising systems. The approach can be viewed as an extension of the mean field approach, while the variational ansatz is represented by more flexible neural networks. To generalize the strategy to many-body physics systems with continuous variables, Ref.~\cite{Wang:2020hji} introduced the Beta distribution layer and wave-net structure for constructing a continuous autoregressive network (CAN), and demonstrated it on the 2D XY model with the thermodynamics of the systems and underlying KT transition well-recognized (see the right panel of Fig.~\ref{probcond}). Interestingly, by inspecting the output of the trained CAN on given XY spin configurations in the topological phase, it was found that the predicted conditional probabilities map matched automatically with the vorticity map of the configuration, implying that perception of the topological vortex structure as the underlying effective degrees of freedom of the physical system naturally emerge from the trained network. 

(4) {\it{Application of Normalizing Flow}} :
As another popular ML deep generative method, normalizing flow (NF) is currently under extensive exploration in the context of lattice QFT for phase transition studies. Being similar to the above-mentioned variational generative approach, NF uses neural networks to give a variational distribution which is implicitly constructed via an affine bijective function transformation represented by the network and exerted on a naive prior noise variable ($z\sim p(z)$) to generate the desired field configuration ($\phi$): $\phi=g_{\theta}(z)$. Through the change of variable, the corresponding variational distribution under such NF model is $p_{\theta}(\phi)=p(z)|\det(\frac{\partial z}{\partial\phi})|$. With Real NVP structure to simplify the Jacobian determinant evaluation in the variational calculation, such flow-based models have been demonstrated for lots of different QFT systems~\ucite{Albergo:2019eim, Kanwar:2020xzo}, including also the QCD system recently~\ucite{Abbott:2022hkm}. To alleviate the multimode sampling difficulty encountered, the Fourier transformation is introduced into the flow generative models and shown promising in generating Feynman paths for quantum systems~\ucite{Chen:2022ytr}, which can be further applied in phase transition exploration for different systems.

{\textbf{Summary}} In this article, we reviewed a couple of topics for phase transition studies leveraged by machine learning methods. Particular emphasis is placed on phase transitions within condensed matter, nuclear liquid-gas phase transition, nuclear temperature determination, and the identification of QCD transition as well as the inference of QCD EoS from high energy heavy-ion collisions. We provide a succinct introduction to several ML techniques deployed in these phase transition studies, demonstrating through these applications the capacity and invaluable insights ML offers to intricate physics exploration. Overall, discriminative learning algorithms like CNN have been shown promising in identifying phase transitions, and Bayesian techniques provide a systematic way to unveil the thermodynamics related to phase transitions, while unsupervised deep generative AI models would facilitate the non-perturbative exploration of phase transition study.

\vspace{1.cm}

{\it Acknowledgements}: This work is partially supported by the National Natural Science Foundation of China under Contracts No. $11890710$, $11890714$, and $12147101$, and the BMBF funded KISS consortium (05D23RI1) in the ErUM-Data action plan.

\end{document}